\documentclass[twocolumn, showpacs,preprintnumbers,amsmath,amssymb]{revtex4}

\usepackage{graphicx}
\usepackage{dcolumn}
\usepackage{bm}

\begin{document}

\title{ Pressure-induced quantum critical point in a heavily hydrogen-doped iron-based superconductor LaFeAsO   }

\author{  Masayoshi Takeuchi$^{1}$, Naoki Fujiwara$^{1}$\footnote{Email: naoki@fujiwara.h.kyoto-u.ac.jp}, Takanori Kuwayama$^{1}$, Satoshi Nakagawa$^{1}$, Soshi Iimura$^{2}$, Satoru Matsuishi$^{3}$, Youichi Yamakawa$^{4}$, Hiroshi Kontani$^{4}$, and  Hideo Hosono$^{2, 3}$ }


\affiliation{$^1$ Graduate School of Human and Environmental Studies, Kyoto University, Yoshida-Nihonmatsu-cyo, Sakyo-ku, Kyoto 606-8501, Japan\\
$^2$ Institute for Innovative Research, Tokyo Institute of Technology, 4259 Nagatsuda,  Midori-ku, Yokohama 226-8503, Japan\\
$^3$ Materials Research Center for Element Strategy, Tokyo Institute of Technology, 4259 Nagatsuda, Midori-ku, Yokohama 226-8503, Japan\\
$^4$ Department of Physics, Nagoya University, Furo-cho, Nagoya 464-8602, Japan}


\begin{abstract}
{  An iron-based superconductor LaFeAsO$_{1-x}$H$_x$ (0 $\leq x \leq$ 0.6) undergoes two antiferromagnetic (AF) phases upon H doping.  We investigated the second AF phase ($x$=0.6) using NMR techniques under pressure. At pressures up to 2 GPa, the ground state is a spin-density-wave state with a large gap; however, the gap closes at 4.0 GPa, suggesting a pressure-induced quantum critical point. Interestingly, the gapped excitation coexists with gapless magnetic fluctuations at pressures between 2 and 4 GPa. This coexistence is attributable to the lift up of the $d_{xy}$ orbital to the Fermi level, a Lifshitz transition under pressure.
}
\end{abstract}


\maketitle

\section{Introduction}
 The interplay between superconductivity and magnetism  has been the focus of effort to understand the  nature of iron-based systems. A typical phase diagram shows an antiferromagnetic (AF) phase followed by a superconducting (SC) phase with increasing carrier doping or isovalent substitution. Some compounds, such as the 122 system except for CaFe$_2$As$_2$ family \cite{Saha1, Nohara, Saha2}, exhibits the coexistence of both states, suggesting an intimate relationship between the two states. However, the 1111 system exhibits little or no overlap of these states.  A prototypical 1111 family member, LaFeAsO$_{1-x}$H$_x$  ($0 \leq x \leq 0.6$), exhibits unique electronic properties in a heavily electron-doped regime; these properties cannot be found in other iron-based pnictides. Similarly to other iron-based pnictides, an AF phase for nondoped or lightly H-doped samples ($x \lesssim 0.05$) is followed by an SC phase when the doping level is increased; however, the SC phase, with a double-domes structure, expands in a wide range ($0.05 \lesssim x \lesssim 0.5$) \cite{Iimura} and another AF phase develops upon further H doping ($0.5 \lesssim x$) \cite{Fujiwara1, Hiraishi, Sakurai, Fujiwara2, Tamatsukuri}.

 The multiorbital tight-binding model shows that both Fermi surfaces and nesting vectors change upon H doping. In a lightly H-doped regime, the hole pockets at the $\Gamma$ point and at ($\pi$ , $\pi$) are comparable in size with the electron pockets at ($\pi$, 0) or (0, $\pi$), leading to a nesting of $\textit{\textbf{Q}}$=($\pi$, 0). For $x=0.4$, the electron pockets expand and the hole pocket at the $\Gamma$ point almost disappears, leading to a change in the nesting vector from $\textit{\textbf{Q}}$=($\pi$, 0) to $\textit{\textbf{Q}}$=($\pi$, $\pi/3$) \cite{Yamakawa}.  Similar Fermi surfaces were also derived via dynamical mean-field theory plus density-functional theory \cite{Moon}. The change in the nesting vector can cause changes in the wave-vector ($q$)- and frequency ($\omega$)-dependent spin susceptibility $\chi (q, \omega)$, which makes possible the emergence of another AF phase. The emergence is indicative of a Lifshitz transition caused by carrier doping \cite{Fujiwara1, Yamakawa}.

 The $x$-$T$ and $P$-$T$ phase diagrams are presented in Fig. 1. LaFeAsO$_{1-x}$H$_x$ has rather simple phase diagrams because FeAs planes are free from atomic substitution, whereas a variety of SC and AF phases manifests in the case of LaFeAs$_{1-x}$P$_x$O \cite{Lai, Mukuda}, leading to complex phase diagrams. As can be seen from the $x$-$T$ phase diagrams, the second AF phase is adjacent to the SC double domes at ambient pressure, and some of the features of a quantum critical point (QCP) manifest at the phase boundary \cite{Sakurai}. According to the resistivity measurements under pressure, the SC double domes become a single dome at 3 GPa, and the single dome shifts to a lower doping regime as a whole with increasing pressure \cite{Takahashi}. The second AF phase is strongly suppressed at 3.0 GPa, and a "bare" AF QCP manifests at $x\sim 0.55$ (the blue closed circle in Fig. 1) \cite{Fujiwara2}. Critical spin fluctuations develop around the "bare" doping-induced AF QCP as demonstrated by the relaxation rate of $^{75}$As divided by temperature ($1/T_1T$), a measure of low-energy spin fluctuations \cite{Fujiwara2}. The color maps of  $1/T_1T$, (a) and (b) in Fig. 1, demonstrate that the critical spin fluctuations are not directly involved in raising $T_c$, because the highest $T_c$ are realized in the intermediate doping regime ($x\sim0.2-0.3$) where spin fluctuations are almost absent \cite{Kawaguchi}.

\begin{figure}
\includegraphics[scale=0.5]{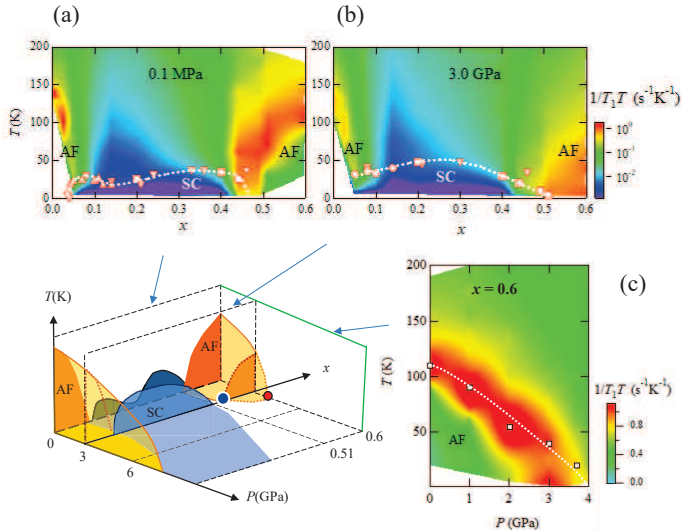}
\caption{\label{fig:epsart} { Electronic phase diagram and color maps for $^{75}$As $1/T_1T$ on LaFeAsO$_{1-x}$H$_x$ and LaFeAsO$_{1-x}$F$_x$. The blue and red closed circles represent doping- and pressure-induced antiferromagnetic (AF) quantum critical points (QCP), respectively. In the color maps (a) and (b), circles and squares represent $T_c$s determined from $1/T_1T$ for F- and H-doped samples, respectively. Upright and inverted triangles represent $T_c$s determined from resistivity for F- and H-doped samples, respectively. In the color map (c) for $x=0.6$,  $T_N$s determined from $1/T_1T$ are plotted as squares, and the dotted curve is a guide to the eye.}}
\end{figure}

 Structural changes in FeAs tetrahedra under pressure have been investigated using X-ray analysis \cite{Kobayashi}.  When the doping level approaches the AF QCP from $x=0.6$ at 3.0 GPa, the As-Fe-As bond angle is estimated to change from 109.3$^{\circ}$ to 109.7$^{\circ}$, and the As height from 1.39{$\AA$} to 1.38{$\AA$}, respectively. In the present work, we measured NMR up to 4.0 GPa for $x$=0.6 and we found that the AF phase disappears at $\sim$ 4.0 GPa. One of the main results is presented in the color map (c) in Fig. 1. We investigate the critical fluctuations near the pressure-induced QCP (the red closed circle in Fig. 1) based on the Fermi surfaces derived from the first-principles analysis.

\begin{figure}
\includegraphics[scale=0.5]{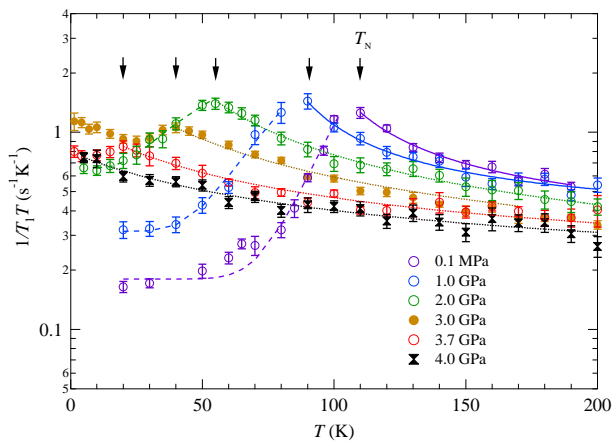}
\caption{\label{fig:epsart} Temperature dependence of $1/T_1T$ for $^{75}$As. Allows represents the AF transition temperatures ($T_N$). The data above and below $T_N$ are fitted to Eq. (1)  and (2), respectively.    }
\end{figure}

\section{Experimental Techniques}
The $^{75}$As-NMR measurements for powder samples were performed using a conventional coherent-pulsed NMR spectrometer. The relaxation rate ($1/T_1$) was measured using a conventional saturation-recovery method. The $^{75}$As-NMR spectra exhibited typical double edges in the field-swept spectra due to the quadrupole interaction; $1/T_1$ was measured at the lower-field edge, where the FeAs planes are parallel to the applied field. The NMR measurements were performed at pressures up to 4.0 GPa using a NiCrAl-hybrid piston-cylinder pressure cell. The detail of the pressure cell is shown in Ref. \cite{FujiwaraNiCrAl}. We first succeeded in detecting the $^{75}$As-NMR signal of the 1111 system at 4.0 GPa by using this pressure cell under a steady load. For a conventional clamped-type pressure cell, the pressure inside the sample space usually decreases by 10\% after the load is released from an oil press. To avoid this 10\% decrease, we mounted an oil press on the NMR probe and applied the steady load. This allowed us to perform NMR measurements at 4.0 GPa. We used a mixture of Fluorinert FC-70 and FC-77 as the pressure-transmitting medium. A coil wounded around the powder samples and an optical fiber with the Ruby powders glued on the top were inserted into the sample space of the pressure cell. The pressure was monitored through Ruby fluorescence measurements.

\section{Experimental results}
Figure 2 shows $1/T_1T$ for $^{75}$As, providing a measure of low-energy magnetic fluctuations: $1/T_1T \sim \sum_{q} Im \chi (q, \omega_N)/\omega_N$ where $\omega_N$ is an NMR frequency. According to a theoretical investigation regarding two-dimensional AF systems \cite{Moriya}, $1/T_1T$ exhibits a Curie-Weiss upturn toward the Neel temperature ($T_N$) with decreasing temperature, and $1/T_1T$ diverges or adopts a maximum at $T_N$. The data were fitted to the following formula:

\begin {equation}
\frac{1}{T_1T} \sim \frac{C}{T-\theta} + const
\end {equation}
where $C$ is a constant.  In conventional AF systems, $\theta$ is equal to $T_N$. The results of $\theta$ are plotted in Fig. 3(a). At low pressures where $\theta$ is positive, $1/T_1T$ exhibits an activated-type magnetic excitation with a large gap, implying that AF ordering is stable.

The gap $\Delta$ of the AF state (Fig. 3(b)) is estimated using the following formula:
\begin {equation}
 \frac{1}{T_1T} \sim e^{-\Delta/T}+const .
\end {equation} The AF phase accompanied by the gap is indicative of a spin-density-wave (SDW) state originating from the nesting between the electron pockets. The constant term arises from the contribution of the Fermi surfaces not involved in the nesting. The Fermi surfaces and the nesting condition are described in detail in the section IV (see Fig. 5).  The value of $\Delta$  was uniquely determined at low pressures below 2 GPa; however, it was hard to determine them uniquely at high pressures because Curie-Weiss (CW) behavior manifests both below and above $T_N$. At high pressures, Curie-Weiss behavior coexists with  the SDW state, and $\theta$ also becomes negative simultaneously. These results suggest that a simple SDW scenario breaks at high pressures above $\sim$ 2 GPa. The maximum of $1/T_1T$ disappears at 4.0 GPa, and therefore, the gap almost closes, as shown in Fig. 3(c). Critical fluctuations around the QCP become predominant throughout the $T$ range at 4.0 GPa.

\begin{figure}
\includegraphics[scale=0.6]{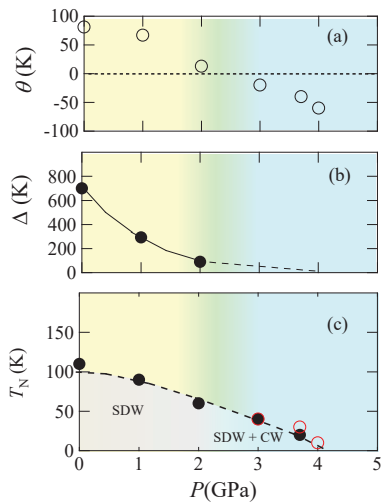}
\caption{\label{fig:wide} Pressure dependence of (a) the Weiss temperature defined in Eq. (1), (b) the excitation gap defined in Eq. (2), and (c) the phase diagram for $x=0.6$. $T_N$ determined from Fig. 2 is shown as black closed circles.The red open circles represent $T_N$ determined from the linewidth shown in Fig. 4. The spin-density-wave (SDW) state is stable for the region colored in yellow ($\theta > 0$). }
\end{figure}

\begin{figure}
\includegraphics[scale=0.5]{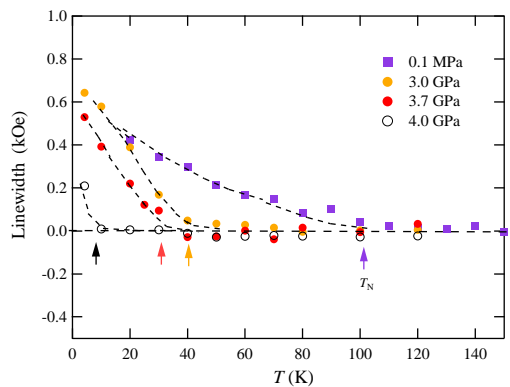}
\caption{\label{fig:wide} Temperature dependence of the $^1$H linewidth at pressures above 3.0 GPa. The linewidth at ambient pressure is also shown for comparison.  }
\end{figure}

Figure 4 shows the $T$ dependence of the $^{1}$H linewidth above 3.0 GPa together with that at ambient pressure \cite{Fujiwara2}. Unlike $^{75}$As NMR spectra, the single $^{1}$H signal broadens in the ordered state. The onset of the broadening, shown by arrows, agrees well with $T_N$ determined from $1/T_1T$. The linewidth broadened below 10 K even at 4.0 GPa, despite the maximum of $1/T_1T$ being absent. This suggests that the QCP is a little higher than 4.0 GPa.

\begin{figure}
\includegraphics[scale=0.3]{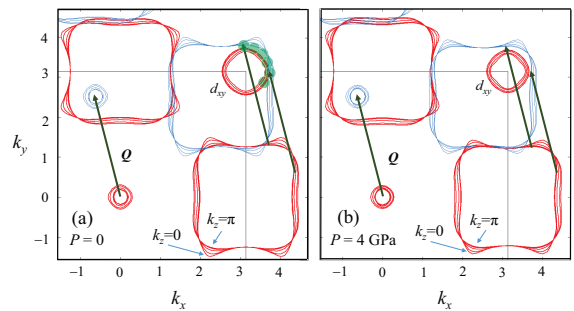}
\caption{\label{fig:wide} Two-dimensional Fermi surfaces for various $k_z$ obtained from the first principle calculations at (a) ambient pressure and (b) 4.0 GPa. Arrows represent the nesting vector $\textit{\textbf{Q}}=(\pi /5, 4 \pi /5)$ connecting the original Fermi surfaces in red and the translated Fermi surfaces in blue. Green circles represent the area of $|\Delta \textbf{\emph{Q}}|\leq $ 0.12. }
\end{figure}

\section{Discussion}
Unlike a simple SDW scenario, gapped and gapless excitations were found to coexist below $T_N$ at pressures between 2 and 4 GPa. This phenomenon indicates topological changes in the Fermi surfaces under pressure. Experimentally, direct observations of the Fermi surfaces have not been made thus far, although photoemission experiments were performed for the powder samples at ambient pressure \cite{Shimojima}. Instead, the Fermi surfaces were theoretically obtained using the first-principles analysis. Figures 5(a) and 5(b) show the Fermi surfaces obtained at ambient pressure and 4.0 GPa, respectively. The crystallographic parameters for $x=0.6$ were estimated from the experimental data for up to $x=0.51$ \cite{Kobayashi}. The nesting vector of $\textit{\textbf{Q}}=(\pi /5, 4 \pi /5)$ results in the maximum overlap of Fermi surfaces. Therefore, the gap below $T_N$ would arise from the SDW state with $\textit{\textbf{Q}}=(\pi /5, 4 \pi /5)$.  However, the CW behavior below $T_N$ can hardly be explained from the pressure dependence of Fermi surfaces alone (see Figs. 5(a) and 5(b)).  The CW behavior below $T_N$ would originate from $\chi(\textbf{\emph{Q}}+\Delta \textbf{\emph{Q}})$ that increases at high pressures but disappears at low pressures. To explain such behavior of  $\chi(\textbf{\emph{Q}}+\Delta \textbf{\emph{Q}})$, the hole pocket of the $d_{xy}$ orbital inside the translated electron pocket would play a key role. In general, a region of $k$ space ($ \sim \Delta/v_F$ where $v_F$ is the Fermi velocity) is involved in the SDW formation. The order parameter $\Delta$  is large at low pressures, and a wide region of $k$ space is involved in the SDW formation.  Therefore, the hole pocket of the $d_{xy}$ orbital could partially or entirely disappear below $T_N$.  The value of $\Delta/v_F$ is estimated to be $\sim $0.12 assuming  $m^*/m=$ 6 where $m^*$ and $m$ are effective mass and band mass, respectively. Green circles in Fig. 5 (a) represent the area of $|\Delta \textbf{\emph{Q}}|\leq $ 0.12.  In various non-doped iron-based compounds, $m^*/m$ is 3 to 6 \cite{Kotliar}. The circles in Fig. 5(a) would become larger and cover a wider area when the mass enhancement is expected \cite{Shishido, Walmsley}. The $\Delta$ being small at high pressures, $|\Delta \textbf{\emph{Q}}|$ becomes extremely small and the entire hole pocket would remain in $k$ space,  resulting in the CW behavior below $T_N$. The presence of the CW behavior suggests the lift up of the $d_{xy}$ orbital to the Fermi level. The lift up of the $d_{xy}$ orbital under pressure has also been predicted for FeSe based on the first-principles analysis \cite{YamakawaFeSe}. The emergence of AF ordering on FeSe under pressure \cite{Sun} can be explained by the lift up of the $d_{xy}$ orbital to the Fermi level and the enhancement of AF fluctuations \cite{YamakawaFeSe}.

The critical fluctuations observed in $1/T_1T$ at 4.0 GPa is also seen near the doping-induced QCP (the blue circle in Fig. 1)\cite{Fujiwara2}. Therefore, it may be concluded that they are inherent in the SDW phase and develop around the second AF phase boundary. However, such fluctuations are not associated with superconductivity as seen from the color maps (a) and (b) in Fig. 1. Moreover, the SC and second AF phases become exclusive with each other as pressure is increased. The present results seem not to contradict with the mechanism via orbital fluctuations \cite{Onari, Kon}. However, magnetic fluctuations are paid attention as a major candidate for the pairing mechanism. In the mechanism via magnetic fluctuations, $d$-wave symmetry is preferred in a heavily doped regime.  A crossover from $s$-wave to $d$-wave symmetries \cite{Fernandes} may be expected in a wide doping range. However, even if a crossover is expected in the present system at ambient pressure, the question of how two symmetries merge under high pressure remains an open problem. In another possibility, high-energy fluctuations may be effective in the pairing. However, the question of why $T_c$ is detectable from the low-energy probe $1/T_1T$ remains to be answered. In any case, clarifying the difference between the electronic states around two QCPs, namely pressure- and doping-induced QCPs, shed lights on the determination of the pairing mechanism.

\section{Conclusion}
We studied the second AF phase of 60\%-H doped LaFeAsO under pressure using NMR techniques. The results revealed unique multi-orbital electronic properties in a heavily electron-doped regime which cannot be found in other iron-based pnictides. Namely, at low pressures below 2 GPa, the ground state is a stable SDW state with a large $\Delta$. The SDW phase almost disappears at 4.0 GPa, causing the "bare" pressure-induced QCP to emerge. Intriguingly, at pressures between 2 and 4 GP, the SDW phase has a small $\Delta$, and gapped excitation coexists with CW behavior. The Fermi surfaces derived from the first-principles analysis suggest that the CW behavior is attributed to the lift up of the $d_{xy}$ orbital to the Fermi level, a Lifshitz transition under pressure.

\begin{acknowledgements}
The NMR measurements under pressure were performed in Kyoto University under the support of JSPS KAKENHI (grant number JP18H01181) and Mitsubishi Foundation, and the first-principles analysis was performed in Nagoya University. We thank Hiroki Takahashi for discussions about resistivity measurements under high pressures.
\end{acknowledgements}

\ \\

\end{document}